\documentclass[final,english,prl,superscriptaddress,showpacs,rotating,lengthcheck]{revtex4}
\usepackage[T1]{fontenc}
\usepackage[latin9]{inputenc}
\usepackage{verbatim}
\usepackage{amsmath}
\usepackage{graphicx}
\usepackage{amssymb}

\makeatletter
\@ifundefined{textcolor}{}
{%
 \definecolor{BLACK}{gray}{0}
 \definecolor{WHITE}{gray}{1}
 \definecolor{RED}{rgb}{1,0,0}
 \definecolor{GREEN}{rgb}{0,1,0}
 \definecolor{BLUE}{rgb}{0,0,1}
 \definecolor{CYAN}{cmyk}{1,0,0,0}
 \definecolor{MAGENTA}{cmyk}{0,1,0,0}
 \definecolor{YELLOW}{cmyk}{0,0,1,0}
 }

\usepackage{amsfonts}

\usepackage{graphics}
\providecommand{\U}[1]{\protect\rule{.1in}{.1in}}

\makeatother

\usepackage{babel}

\makeatother

\usepackage{babel}

\begin{document}

\title{Perfect state transfers by selective quantum interferences within
complex spin networks}

\author{Gonzalo A. \'{A}lvarez}

\altaffiliation{Present address: Fakult\"{a}t Physik, Universit\"{a}t Dortmund, Otto-Hahn-Strasse 4, D-44221
Dortmund, Germany.}

\affiliation{Facultad de Matem\'{a}tica, Astronom\'{i}a y F\'{i}sica, Universidad Nacional
de C\'{o}rdoba, 5000 C\'{o}rdoba, Argentina.}

\author{Mor Mishkovsky}

\altaffiliation{Present address: Ecole Polytechnique Fédérale de Lausanne, CH-1015 Lausanne, Switzerland.}

\affiliation{Department of Chemical Physics, Weizmann Institute of Sciences, 76100
Rehovot, Israel.}

\author{Ernesto P. Danieli}

\altaffiliation{Present address: ITMC, RWTH Aachen, D-52074 Aachen, Germany.}

\affiliation{Facultad de Matem\'{a}tica, Astronom\'{i}a y F\'{i}sica, Universidad Nacional
de C\'{o}rdoba, 5000 C\'{o}rdoba, Argentina.}

\author{Patricia R. Levstein}

\affiliation{Facultad de Matem\'{a}tica, Astronom\'{i}a y F\'{i}sica, Universidad Nacional
de C\'{o}rdoba, 5000 C\'{o}rdoba, Argentina.}

\author{Horacio M. Pastawski}

\affiliation{Facultad de Matem\'{a}tica, Astronom\'{i}a y F\'{i}sica, Universidad Nacional
de C\'{o}rdoba, 5000 C\'{o}rdoba, Argentina.}

\author{Lucio Frydman}

\altaffiliation{Corresponding author: Lucio.Frydman@weizmann.ac.il}

\affiliation{Department of Chemical Physics, Weizmann Institute of Sciences, 76100
Rehovot, Israel.}


\begin{abstract}
We present a method that implement directional, perfect state transfers
within a branched spin network by exploiting quantum interferences
in the time-domain. That provides a tool to isolate subsystems from
a large and complex one. Directionality is achieved by interrupting
the spin-spin coupled evolution with periods of free Zeeman evolutions,
whose timing is tuned to be commensurate with the relative phases
accrued by specific spin pairs. This leads to a resonant transfer
between the chosen qubits, and to a detuning of all remaining pathways
in the network, using only global manipulations. As the transfer is
perfect when the selected pathway is mediated by $2$ or $3$ spins,
distant state transfers over complex networks can be achieved by successive
recouplings among specific pairs/triads of spins. These effects are
illustrated with a quantum simulator involving $^{13}$C NMR on Leucine's
backbone; a six-spin network. 
\end{abstract}
\maketitle
Quantum state control is essential in quantum information processing.
In particular, the perfect state transfer (PST) between two-level
systems (qubits) is one of the building blocks underlying quantum
communications. 
Spin chains have been proposed as candidates for exploring short-distance
communications \cite{Bose03} and \emph{XY}-type (flip-flop) interactions
tend to be essential to the efficiency of these quantum effects \cite{Pastawski96}.
With such couplings, ideal quantum PSTs have been demonstrated in
two and three spin chains \cite{Laflamme,Christandl04}. Perfect transfers,
however, are still a challenge for larger spin chains --particularly
for complex, branched spin networks. PST proposals to deal with these
cases include engineering the spin-spin coupling intensities and/or
the spin energies \cite{Christandl04,Stolze05,Jafariyadeh08,Gagnbin07},
as well as turning on-and-off the local couplings \cite{Laflamme,Glaser01}.
A general feature of all these proposals is their requirement to access
individual spins; either during the preparation of the quantum {}``hardware\textquotedblright{},
or during the transfer procedure. Either of these conditions is hard
to implement with most present technologies. In fact, owing to the
flexibility offered by active decoupling methods, heteronuclear nuclear
magnetic resonance (NMR) presents one of the few scenarios where PSTs
were demonstrated \cite{Laflamme}. On the other hand, given the scarcity
of heteronuclear qubits that can be simultaneously included in any
one system, homonuclear-based methods could provide a much better
platform for exploring PSTs and/or controlling other aspects of spin
dynamics. Homonuclear NMR realizations of partial state transfers
have actually been demonstrated; both by Mádi \emph{et al.}, who explored
a liquid-state linear chain system \cite{Madi97}, and more recently
by Cory \emph{et al.,} who proposed measuring state transfers in a
solid state sample with a linear coupling topology by relying on time-averaging
techniques that requires simultaneous manipulations of individual
spins \cite{Cory07}. Alternative theoretical proposals by Khaneja
and Glaser \cite{Khaneja02} and by Fitzsimons and co-workers \cite{Fitzsimons}
allow for perfect transfers, by manipulating solely those spins at
the end of linear Ising spin chains. 

Here, we introduce a different method for carrying out PSTs within
a complex homonuclear system, that does not require any manipulation
of individual spins. Instead, we exploit time domain quantum interferences
that arise when spins evolve in the presence of non-commuting \emph{XY}
spin-spin coupling and Zeeman-type longitudinal Hamiltonians. It is
shown that global manipulations with these interactions \cite{Tycko06},
can effectively turn on-and-off specific local couplings within the
network, and achieve a full transfer of quantum information between
distant qubits regardless of the spin network topology. We discuss
some consequences of these ideas within the context of a liquid-state
NMR {}``quantum PST simulator\textquotedblright{}.

\begin{figure}[tbh]
\begin{centering}
\includegraphics[width=3.1038in,height=3.1073in]{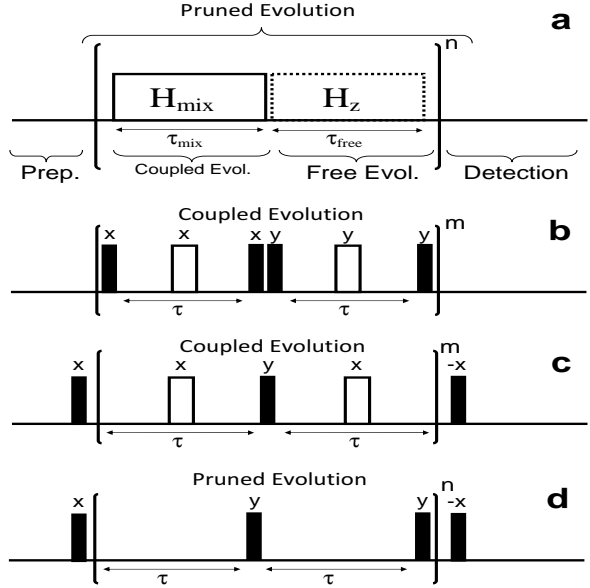} 
\par\end{centering}

\caption{Quantum evolution schemes for achieving perfect state transfers (PSTs)
in homonuclear NMR systems. (a) Sequence delivering PST when $\widehat{\mathcal{H}}_{\mathrm{mix}}=\widehat{\mathcal{H}}_{\mathrm{mix}}^{XY}$
and a suitable number $n$ of repetitions is used. (b, c) NMR mixing
sequences to generate an effective $XY$ Hamiltonian in liquid state
systems, starting from a spin-spin Ising-type interaction \cite{Madi97}.
$m$ is the number of loops defining the resulting $\widehat{\mathcal{H}}_{\mathrm{mix}}$
Hamiltonian. (d) Alternative sequence incorporating quantum evolutions
under the combined effects of $\widehat{\mathcal{H}}_{\mathrm{mix}}$
and $\widehat{\mathcal{H}}_{\mathrm{Z}}$ Hamiltonians, capable of
generating a PST between specific homonuclear offsets based on an
Ising spin-spin interaction. Solid lines represent $\pi/2$ pulses;
hollow boxes are $\pi$ pulses.}

\centering{}\label{Fig_NMRseq} 
\end{figure}

\emph{Qubit-Selective Perfect Transfers in a Network.--} Consider
a spin network subject to the rotating-frame Hamiltonian \begin{equation}
\widehat{\mathcal{H}}=\widehat{\mathcal{H}}_{\mathrm{Z}}+\widehat{\mathcal{H}}_{\mathrm{mix}}=\sum_{i}\hbar\Delta\Omega_{i}\hat{I}_{i}^{z}+\sum_{i\neq j}J_{ij}^{{}}\left(\hat{I}_{i}^{x}\hat{I}_{j}^{x}+\hat{I}_{i}^{y}\hat{I}_{j}^{y}\right).\label{eq:H}\end{equation}
 $\widehat{\mathcal{H}}_{\mathrm{Z}}$ is the longitudinal interaction
of the spins defined by their $\Delta\Omega_{i}$ chemical shifts,
and $\widehat{\mathcal{H}}_{\mathrm{mix}}$ is a suitable mixing Hamiltonian.
We focus in particular on an $XY$ mixing interaction, $\widehat{\mathcal{H}}_{\mathrm{mix}}^{XY}$,
as it has been shown \cite{Christandl04} that this allows PSTs to
occur over 2 or 3 spins by setting every qubit/spin that belong to
the channel to $\left|0\right\rangle =\left|\downarrow\right\rangle _{z}$
and the source spin in a quantum superposition $\alpha\left|0\right\rangle +\beta\left|1\right\rangle $,
where $\left|1\right\rangle =\left|\uparrow\right\rangle _{z}$ .
Assuming that $\Delta_{ij}=\hbar\left\vert \Delta\Omega_{i}^{{}}-\Delta\Omega_{j}^{{}}\right\vert \gg J_{ij}$
is satisfied, then the $\widehat{\mathcal{H}}_{\mathrm{mix}}^{XY}$
term in Eq. (\ref{eq:H}) turns non-secular and a free evolution is
in consequence achieved: spins evolve in isolation from one another.
Conversely, to optimize a transfer of polarization between sites,
the longitudinal interaction needs to be erased --for example by using
radiofrequency (rf)-based methods that refocus the shifts of all spins
\cite{Tocsy}. The system will then evolve with a nearly pure mixing
Hamiltonian $\widehat{\mathcal{H}}_{\mathrm{mix}}$; as this Hamiltonian
connects all possible spin pairs, an initial local excitation will
spread over the full spin network. To avoid this and confine the spin
network topology to a selected, optimized spin transfer pathway, one
needs to effectively disconnect spins that do not belong to this sub-network.
We achieve this by exploiting the chemical shift differences among
the sites involved; specifically, by implementing a sequence whereby
the $\widehat{\mathcal{H}}_{\mathrm{mix}}$ evolution is stroboscopically
interrupted with free evolution periods driven by $\widehat{\mathcal{H}}_{\mathrm{Z}}$
(Fig. \ref{Fig_NMRseq}a). If periods of coupled and free evolutions
are alternated $n$-times,\ and the free evolution time, $\tau_{\mathrm{free}}$,
is adjusted to a common multiple of the inverses of the chemical shift
differences $\Delta_{ij}$ between the spins of the pathway (i.e.
$\tau_{\mathrm{free}}=2\pi n_{ij}\hbar/\left\vert \Delta_{ij}\right\vert ,$
where $n_{ij}$ is a positive integer), only the $i$ and $j$ spins
in the pathway experience a coherent $\widehat{\mathcal{H}}_{\mathrm{mix}}$
--driven buildup. And only between these spins the PST proceeds coherently.
By contrast all other pathways are dephased by the free evolution,
leading to their effective decoupling. %
This decoupling turns more effective by reducing the $\widehat{\mathcal{H}}_{\mathrm{mix}}$
evolution time $\tau_{\mbox{mix}}$ because it increases the detuning
between the selected spins and the remaining ones as can be observed
by the zero-order average Hamiltonian $\left(\tau_{\mbox{free}}\widehat{\mathcal{H}}_{\mathrm{Z}}^{\ne i\ne j}+\tau_{\mbox{mix}}\widehat{\mathcal{H}}_{\mbox{mix}}\right)/\left(\tau_{\mbox{free}}+\tau_{\mbox{mix}}\right)$,
where $\widehat{\mathcal{H}}_{\mathrm{Z}}^{\ne i\ne j}$ is a Zeeman
Hamiltonian for the remaining spins. Moreover, by judiciously turning
on-and-off transfers within any two/three qubits, one can control
by PSTs between arbitrarily chosen pairs of spins even throughout
complex spin network topologies. From a quantum-mechanical standpoint
this can be interpreted as a successive establishment of decoherence-free
subspaces, where the only allowed dynamics occurs within a confined
region of the spins' Hilbert space \cite{Pascazio}. 
Indeed the repetitive interruption of $\widehat{\mathcal{H}}_{\mathrm{mix}}$
with the strong $\widehat{\mathcal{H}}_{\mathrm{Z}}$ interactions
can be shown equivalent to a stroboscopic measurement of local variables
\cite{Alvarez06}. This results in a dynamical Quantum Zeno Effect
\cite{dQZE} that freezes certain portions of the internal quantum
dynamics, and delivers a nearly ideal PSTs. This is by contrast to
standard PST implementations, that rely on concatenated SWAP gates
(like those afforded by selective pulses) to achieve similar goals
\textendash{}but via more demanding, local manipulations \cite{Glaser01}.

In liquid-state homonuclear systems at high fields, $\left\vert \Delta_{ij}\right\vert \gg J_{ij}$,
and the intrinsic $\widehat{\mathcal{H}}_{\mathrm{mix}}^{XY}$ terms
will be truncated. Although not naturally available, such flip-flop
couplings can be reintroduced in an average way, by toggling the usual
high-field Ising $J$-interaction, $\widehat{\mathcal{H}}_{\mathrm{mix}}^{ZZ}=\sum_{i\neq j}J_{ij}^{{}}\hat{I}_{i}^{z}\hat{I}_{j}^{z}$,
into an effective $XY$ Hamiltonian. Rotating the $\{I_{\alpha}\}_{\alpha=x,y,z}$
states at a high rate with respect to the relevant interactions as
schematized in Fig. \ref{Fig_NMRseq}b, transforms the averaged Ising
Hamiltonian, into an effective $\widehat{\mathcal{H}}_{\mathrm{mix}}^{XY}$
\cite{Madi97}. Fig. \ref{Fig_NMRseq}c shows an alternative sequence,
capable of achieving the same Hamiltonian but in an experimentally
more robust manner. This latter scheme still requires appending a
free evolution period $\tau_{\mathrm{free}}$, tuned to the inverse
shift difference $\Delta_{ij}$ between the pair of sites among which
one intends the PST. Instead of inserting these delays explicitly,
we modified the sequence in Fig. \ref{Fig_NMRseq}c to impose an offset
dependence, which truncates all $J_{ij}$ effects except for those
spin-pairs for which the $\tau^{-1}$ inverse interval is a common
multiple of the chemical shift differences $\Delta_{ij}=\hbar\left(\Delta\Omega_{i}-\Delta\Omega_{j}\right)$.
This alternative (Fig. \ref{Fig_NMRseq}d), which was experimentally
the most robust variant assayed, effectively connects coupled spins
in a pathway if they fulfill $\tau=2\pi n_{ij}\hbar/\left\vert \Delta_{ij}\right\vert $,
while dephasing the transfer between all other pairs of spins. Assuming
that $\left\vert \Delta_{ij}\right\vert \gg J_{ij},$ only the targeted
spin pairs undergo an effective PST.

\begin{figure}[tbh]
\begin{centering}
\includegraphics[width=3.3122in,height=2.5944in]{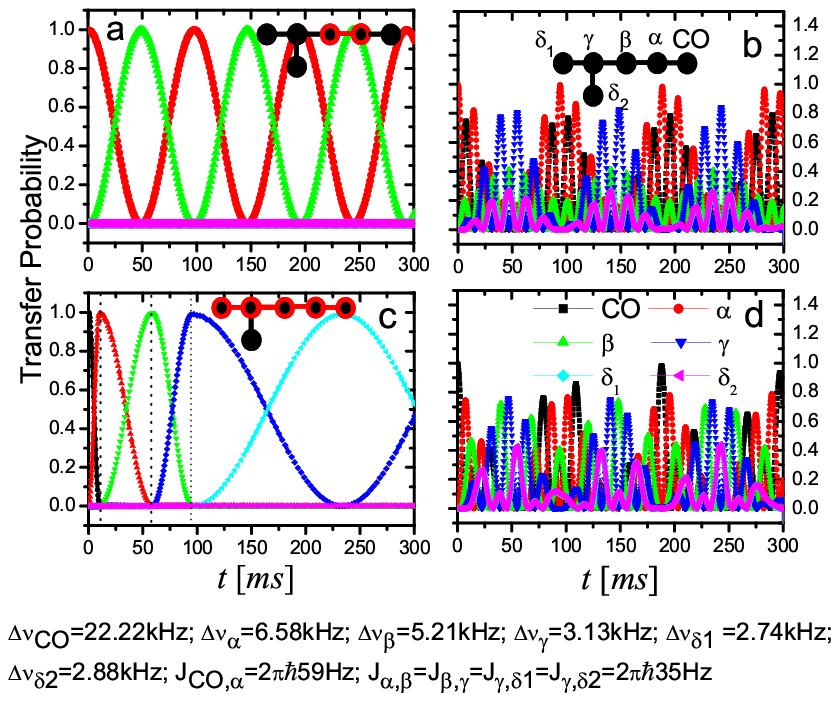} 
\par\end{centering}

\caption{(color online) Numerical simulations for perfect state transfers throughout
different sites of the $^{13}$C L-leucine molecule backbone. $\widehat{\mathcal{H}}_{\mathrm{mix}}$
was assumed a perfect $\widehat{\mathcal{H}}_{\mathrm{mix}}^{XY}$,
and $\widehat{\mathcal{H}}_{\mathrm{free}}^{{}}=\widehat{\mathcal{H}}_{\mathrm{Z}}^{{}}$
was selected with $\tau_{\mathrm{mix}}=0.3\operatorname{ms}\ll2\pi\hbar/J$.
(a) PST between the spin sites $\alpha$ and $\beta$ by selecting
$\tau_{\mathrm{free}}=2\pi\hbar/\Delta_{\alpha\beta}$. (b) Counterpart
evolution starting also from site $\alpha$, but acted upon by a pure
$\widehat{\mathcal{H}}_{\mathrm{mix}}^{XY}$ evolution. (c) Step-wise
PST starting from an initial excitation on the CO spin and concluding
in the $\delta_{1}$ spin. $\tau_{\mathrm{free}}=2\pi\hbar/\Delta_{ij}$
was chosen to effectively reduce the system to a series of two-spin
transfers, proceeding throughout the backbone. (d) Counterpart evolution
expected for all spins in the system starting from an initial CO site
excitation, but proceeding under the action of an unfiltered $\widehat{\mathcal{H}}_{\mathrm{mix}}^{{}}=\widehat{\mathcal{H}}_{\mathrm{mix}}^{XY}$
coupled evolution. Bottom: Scheme of the spin network and their parameters
obtained from NMR experiments.}

\centering{}\label{Fig_graphteo} 
\end{figure}

\emph{Numerical simulations.--} The concepts just mentioned were numerically
tested to explore, their usefulness for performing PSTs within an
ideal spin network. Specifically, we sought to inquire into the efficiency
of the truncated $XY$ mixing Hamiltonian to support long-range transfers
between non-neighboring spins; both directly, as well as through numerous
pair-wise stop-overs involving intermediate spins. For concreteness,
we focused on the six-spin system of the L-leucine's $^{13}$C backbone;
whose chemical shifts $\Delta\nu_{i}=\Delta\Omega_{i}/2\pi$ and $J$-couplings,
obtained experimentally by NMR, are listed in Fig. \ref{Fig_graphteo}.
Numerical simulations of sequence \ref{Fig_NMRseq}a with $\widehat{\mathcal{H}}_{\mathrm{mix}}^{{}}=\widehat{\mathcal{H}}_{\mathrm{mix}}^{XY},$
$\widehat{\mathcal{H}}_{\mathrm{free}}^{{}}=\widehat{\mathcal{H}}_{\mathrm{Z}}^{{}}$
are shown in Figs. \ref{Fig_graphteo}a-d. These plots describe the
fidelity of PSTs solely as a function of the targeted qubits' $\left|\uparrow\right\rangle _{z}$-probability.
Such description is made possible thanks to the fact that the total
$\sum_{i}\hat{I}_{i}^{z}$ angular momentum is a constant of motion:
The transfer of a qubit state on $i$ throughout a network can thus
be gauged if, given an initial state where $\left|i\right\rangle =\left|\uparrow\right\rangle _{z}$
and all remaining spins are $\left|\downarrow\right\rangle _{z}$,
the latter's evolution in time is followed \cite{Christandl04}.
The top and bottom panels of Fig. \ref{Fig_graphteo} focus on such
description assuming two different PST targets, whereas its left/right-hand
sides compare the differences between selectively-transferred and
normal $\widehat{\mathcal{H}}_{\mathrm{mix}}^{XY}$ pathways. Figs.
\ref{Fig_graphteo}a-b illustrate the behavior of the spin up probabilities
that starting in site C$_{\alpha}$, undergoes a fully spin-spin coupled
or a PST-selective $XY$ evolution between the sites C$_{\alpha}$
and C$_{\beta}$. The excellent selectivity of the latter choice is
evidenced by the long-term, coherent nature of the oscillations. Fig.
\ref{Fig_graphteo}c illustrates a different aspect of the PST, whereby
the initial excitation is localized at the CO carbonyl site and, by
successively selecting a train of suitable conditions $\tau_{\mathrm{free}}=2\pi\hbar/\Delta_{ij}$,
this is subsequently passed through all the sites in the main molecular
chain until reaching the end C$_{\delta1}$ site. In other words,
by selecting $\tau_{\mathrm{free}}=2\pi\hbar/\Delta_{\mathrm{CO},\alpha}$
one can do a PST to the C$_{\alpha}$ site; once this state transfer
is maximized one can then set $\tau_{\mathrm{free}}=2\pi\hbar/\Delta_{\alpha,\beta}$
and transfer it to site C$_{\beta}$, and onwards with successive
steps until the initial state has been transferred to the $\delta_{1}$
site. Although relatively time consuming, this step-by-step transfer
is realized with high efficiency and without accessing $\delta_{2}$,
owing to the method's high selectivity. This is in stark contrast
with the very complex behavior observed when allowing the spin system
to evolve under a pure $\widehat{\mathcal{H}}_{\mathrm{mix}}^{XY}$
evolution (Fig. \ref{Fig_graphteo}d). Similar curves are obtained
when the initial excitation is on the other spins. PST between arbitrary
sites in a branched network is thus achieved, without having to address
the individual qubits selectively.

\begin{figure}[tbh]
\begin{centering}
\includegraphics[scale=0.27,angle=270]{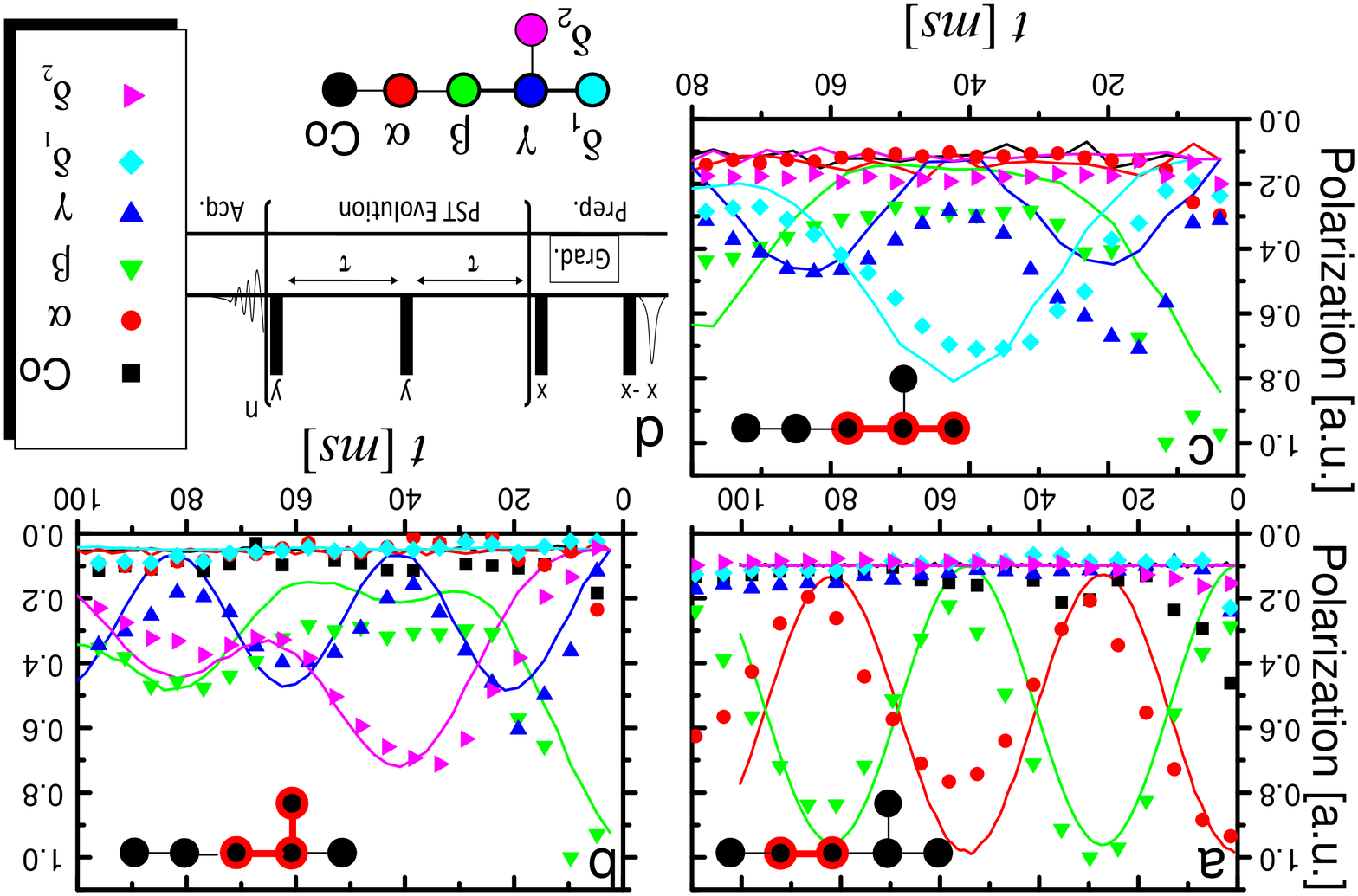} 
\par\end{centering}

\caption{(color online) NMR experiments (points) and numerical expectations
(lines) for selective state transfers implemented for different situations
by using the NMR pulse sequence illustrated in the lower right corner
--derived in turn from Fig. \ref{Fig_NMRseq}d. (a) An initial excitation
on $\alpha$ is transferred to the $\beta$ spin. (b) An initial C$_{\beta}$
excitation is optimally transferred to $\delta_{2}$ via the $\beta-\gamma-\delta_{2}$
pathway. (c) An initial C$_{\beta}$ excitation is selectively transferred
to $\delta_{1}$ via the $\beta-\gamma-\delta_{1}$ pathway.}

\centering{}\label{Fig_exp} 
\end{figure}

\emph{Experiments.--} The new PST approach and its associated features
were tested using liquid state $^{13}$C NMR as a sort of {}``quantum
simulator\textquotedblright{}. These experiments were carried out
at 298 K and 11.7 T, using U-$^{15}$N,$^{13}$C-Leu-FMOC dissolved
in CDCl$_{3}$ as test case. The measured chemical shifts (in kHz
referenced to $\delta_{\mathrm{TMS}}$ = 0), the $J$-couplings, as
well as the spin-coupling topology, were summarized in Fig. \ref{Fig_graphteo}.
Figure \ref{Fig_exp} schematizes the actual pulse sequence assayed
on this test system, together with comparisons between polarization
evolutions observed for different initial conditions and different
pathways as building blocks for piecewise transfers; also shown are
calculations of the expected behavior. To better gauge this comparison
we prepared an initial, local excitation, applying a selective $\pi/2$
pulse on the source site, which turned the initial magnetization of
the selected spin to a perpendicular axis of the static magnetic field.
Then, a non-selective $\pi/2$ pulse restores back to the $z$ axis
the source spin, while all the other spin magnetizations turn transverse
to the static field. These magnetizations were then promptly dephased
by applying a field gradient pulse. This source of polarization, which
is in a single site, is then rotated to the $-y$ axis by a non-selective
pulse, and followed by a mixing sequence akin to that in Fig. \ref{Fig_NMRseq}d.
The transferred polarizations were observed at times $t=2n\tau$.
The rf carrier frequency was set on-resonance with C$_{\beta}$; all
polarizations thus appeared evolving in a rotating frame that precessed
with frequency $\Delta\Omega_{\beta}$. While these examples illustrate
the good PST performance for longitudinal polarizations transfers
which are the equivalent of an excitation transfer probabilities,
a complete performance test within NMR ensemble quantum computation
would require the incorporation of pseudo-pure states into this kinds
of manipulations \cite{pseudo00}.

The main effects of the selective PST are observed very well --even
if some undesired residual magnetization from non-source spins survived
the initial purging process, and despite the limited performance that
our non-selective pulses could achieve given the maximal 19 kHz radiofrequency
fields achievable in our system. In Fig. \ref{Fig_exp}a $\tau$ was
set as $0.71\operatorname{ms}\simeq2\pi\hbar/\left\vert \Delta_{\alpha\beta}\right\vert $
producing a polarization transfer between sites C$_{\alpha}$ and
C$_{\beta}$, while effectively decoupling the remaining $^{13}$C-$^{13}$C
interactions. Figs. \ref{Fig_exp} b-c show additional examples of
pairwise PSTs. Values of $\tau=3.88\operatorname{ms}=8\times2\pi\hbar/\left\vert \Delta_{\beta\gamma}\right\vert \simeq2\pi\hbar/\left\vert \Delta_{\gamma,\delta2}\right\vert $
favored the selective transfer from site C$_{\beta}$ to carbon C$_{\gamma}$
and on to site C$_{\delta2}$ (Fig. \ref{Fig_exp}b); choosing $\delta_{1}$
instead of $\delta_{2}$, $\tau=4.81\operatorname{ms},$ optimized
a similar transfer but for the C$_{\beta}$$\rightarrow$C$_{\delta1}$
case (Fig. \ref{Fig_exp}c). Though the chosen times $\tau$ were
not perfectly commensurate with their ideal values, the functionality
of the method is evident. And the overall agreement with numerical
simulations that consider the entire sequence (lines in Fig. \ref{Fig_exp})
is excellent.

%


\emph{Conclusions.--} A method for achieving a directional, perfect
state transfer within a branched spin network without individual manipulation
of the spins, has been proposed and demonstrated. The method requires
knowledge of system parameters such as the chemical shifts, but does
not need selective qubit manipulations. As a result the method is
general and independent of the system size in contrast to methods
based on selective manipulations where the number of individual manipulations
grows roughly exponentially with the system size \cite{Glaser01}.
Selectively information transfers among proximal or distal qubits
regardless of the spin network topology was then demonstrated, by
engineering a Hamiltonian that exploits the two non-commuting contributions
to the system's evolution: one involving a pure $XY$ interaction,
and the other a pure Zeeman evolution. Alternatives for generating
such selective $XY$ interactions by means of manipulating Ising couplings,
were also shown, acting by establishing decoherence-free subspaces
where complete PSTs between the targeted spins can occur, while effectively
{}``pruning\textquotedblright{} away those qubits where no excitation
transfer is required. While the experimental performance of the approach
was illustrated with PSTs of longitudinal polarizations, the method
can, in general, provide a selective, specific transfer pathway in
a system of many interacting spins for arbitrary initial states for
the source spin.%
{} Moreover, by dividing a complex spin network into smaller sub-ensembles,
this approach provides an excellent starting point for performing
other large quantum systems using solely global manipulations that
exploit the time-domain quantum interferences. Some of these will
be discussed in further detail in upcoming studies.


\begin{acknowledgments}
This work was made possible by a Fundación Antorchas -- Weizmann Institute
Cooperation (grant \#2530), by CONICET, by a Helen and Kimmel Award
for Innovative Investigations, as well as by the generosity of the
Perlman Family Foundation. GAA and EPD thank the Alexander von Humboldt
Foundation for Research Scientist Fellowships. \end{acknowledgments}

\end{document}